\documentclass[12pt]{article}
\usepackage[colorlinks=true,linkcolor=blue, linktoc=page, urlcolor=blue,citecolor=magenta]{hyperref}
\usepackage{amsmath,amscd}
\usepackage{amsfonts}
\usepackage{scalerel}
\usepackage{amssymb}
\usepackage{graphicx,shortvrb}
\usepackage{xcolor}
\usepackage{cancel}
\usepackage{stmaryrd}
\usepackage[utf8]{inputenc}
\usepackage{mdframed}
\usepackage{cite}

\newcommand{\os}[2]{{\overset{\,\scalebox{0.5}{($#2$)}}{#1}}{}}

\def\calc{{\cal C}}

\def\calo{{\cal O}}

\def\calt{{\cal T}}

\def\cale{{\cal E}}

\def\d{\delta}

\def\D{\Delta}

\def\l{\lambda}

\def\m{\mu}
\def\n{\nu}
\def\o{\omega}
\def\O{\Omega}

\def\r{\rho}

\def\s{\sigma}
\def\S{\Sigma}

\def\oa{\os{A}{-1}}
\def\oaa{\os{A}{-2}}
\def\dm{{\dot{\mu}}}
\def\dn{{\dot{\nu}}}
\def\ds{{\dot{\sigma}}}

\def\dr{{\dot{\rho}}}

\def\eg{{\reflectbox{\rotatebox[origin=c]{180}{$\mathbb{L}$}}}}
\def\maxcon{{{\eg}}}

\def\maxnab{{\nabla\!\!\!\!\raisebox{0.161em}{\scaleobj{0.75}{\nabla}}}}

\def\GN{{G_\mathrm{N}}}

\def\ri{\mathrm{i}}

\begin{document}

\begin{titlepage}
 \vskip 1.8 cm

\begin{center}{\huge \bf Oddity in nonrelativistic, strong gravity}\\

\end{center}
\vskip .3cm
\vskip 1.5cm

\centerline{\large {{\bf Mert Ergen$^1$, Efe Hamamc$\i^1$\ and  Dieter Van den Bleeken$^{1,2}$}}}

\vskip 1.0cm

\begin{center}
	1) Primary address\\
	Physics Department, Boğaziçi University\\
	34342 Bebek / Istanbul, TURKEY
	
	\vskip 1cm
	
	2) Secondary address:\\
	Institute for Theoretical Physics, KU Leuven\\
	3001 Leuven, Belgium
	
	\vskip 1cm
	
\texttt{mert.ergen, efe.hamamci, dieter.van\, @boun.edu.tr}
\end{center}
\vskip 1.3cm \centerline{\bf Abstract} \vskip 0.2cm \noindent We consider the presence of odd powers of the speed of light $c$ in the covariant nonrelativistic expansion of General Relativity (GR). The term of order $c$ in the relativistic metric is a vector potential that contributes at leading order in this expansion and describes strong gravitational effects outside the (post-)Newtonian regime. The nonrelativistic theory of the leading order potentials contains the full non-linear dynamics of the stationary sector of GR.

\end{titlepage}
\tableofcontents
\section{Introduction}
The covariant nonrelativistic approximation to GR as introduced by Dautcourt \cite{Dautcourt:1996pm} has recently been revisited \cite{Tichy:2011te, VandenBleeken:2017rij, Hansen:2020, Hansen:2018ofj, Cariglia:2018hyr, VandenBleeken:2019gqa, Hansen:2019vqf, Hansen:2019svu}. The approximation amounts to an expansion of the relativistic metric in inverse powers of the speed of light $c$ and for this reason we will refer to it as the large $c$ expansion. Originally \cite{Dautcourt:1996pm, Tichy:2011te} the aim was simply to provide a manifestly coordinate invariant version of the post-Newtonian expansion. But in \cite{VandenBleeken:2017rij} it was pointed out that the large $c$ expansion naturally allows the inclusion of strong gravitational effects which are not captured\footnote{To be precise, the strong gravitational effects present at leading order in the large $c$ expansion can not be fully reproduced at any finite order of the post-Newtonian expansion.} in the post-Newtonian setup. That the large $c$ expansion extends the post-Newtonian expansion is no surprise, since the latter is not only a nonrelativistic, but also a weak field expansion \cite{Poisson:2014}. This feature of effectively describing some strong gravitational effects suggests the large $c$ expansion could have interesting phenomenological applications. This remains largely unexplored, but we make a few initial observations in section \ref{pheno}. Still this approach has already led to some interesting theoretical progress, such as the formulation of a variational principle for Newtonian gravity \cite{Hansen:2018ofj} and a better understanding of the geometry underlying nonrelativistic gravity \cite{VandenBleeken:2017rij, Hansen:2020}. Furthermore it has some conceptual value, in that it clearly separates relativistic effects from strong gravity effects \cite{Hansen:2019vqf}. The work on the large $c$ expansion, which by construction provides a nonrelativistic gravity theory descendant from GR, is also of value to the recent explorations of more general theories of nonrelativistic gravity. In that context we can mention for example recent work on 3d nonrelativistic gravity \cite{Papageorgiou:2009zc,Andringa:2013mma, Bergshoeff:2015uaa, Bergshoeff:2016lwr, Hartong:2016yrf, Hartong:2017bwq, Bergshoeff:2017dqq, Matulich:2019cdo,Ozdemir:2019orp,Aviles:2019xed,Chernyavsky:2019hyp, Penafiel:2019czp, Ozdemir:2019tby,Concha:2019dqs,Concha:2019lhn,Concha:2019mxx, Bergshoeff:2020fiz}, Lifshitz Holography \cite{Christensen:2013lma,Christensen:2013rfa,Hartong:2014oma,Hartong:2015zia}, nonrelativistic string theory \cite{Gomis:2000bd,Andringa:2012uz,Harmark:2017rpg,Kluson:2018egd,Kluson:2018grx,Harmark:2018cdl,Bergshoeff:2018yvt,Bergshoeff:2018vfn,Gomis:2019zyu,Bergshoeff:2019pij,Gallegos:2019icg,Harmark:2019upf,Kluson:2019ifd,Kluson:2019xuo,Blair:2019qwi,Kluson:2020aoq} and condensed matter and fluid mechanics applications \cite{Son:2013rqa,Jensen:2014aia,Geracie:2014nka,Jensen:2014ama,Geracie:2015xfa,Geracie:2015dea,deBoer:2017abi,deBoer:2017ing,Armas:2019gnb,Bagchi:2019clu}. In parallel the symmetries underlying such theories have been further investigated \cite{Andringa:2010it,Afshar:2015aku,Figueroa-OFarrill:2017sfs,Figueroa-OFarrill:2018ilb,Bergshoeff:2019ctr,Gomis:2019fdh,Figueroa-OFarrill:2019sex,deAzcarraga:2019mdn,Abedini:2019voz,Gomis:2019sqv}. 

Previous work on the large $c$ expansion has focused on an expansion of the metric in inverse powers of $c^2$. This is a self-consistent assumption in the gravitational sector since gravitational potentials of even power in $c$ can not source potentials of odd power in $c$. Under an additional weak field assumption, together with consideration of some physical constraints on energy-momentum and an appropriate coordinate choice, one can show that odd terms in the relativistic metric can appear only at subleading order \cite{Dautcourt:1996pm, Poisson:2014}. In case one does {\it not} make the weak field assumption the presence of terms with an odd power of $c$ has to date remained unexplored and in this work we initiate its study. This is motivated by the observation that energy-momentum that sources the strong time dilation potential -- the twistless torsion of \cite{VandenBleeken:2017rij} -- will also source the leading order odd term in the metric when it is dynamic \cite{Hansen:2020, VandenBleekenx}. 

The main result of this paper is the set of equations presented in table \ref{eqtab}. These are the Einstein and conservation equations, at leading order in an expansion of inverse powers of $c$, with the only assumption made that the relativistic metric and the energy-momentum tensor are of the form $g_{\m\n}\sim c^2\tau_\m\tau_\nu+\calo(c)\,,$  $T_{\m\n}\sim c^{6}\tau_\m\tau_\n+\calo(c^{5})$. The physical fields at the leading order are a scalar potential $\Psi$, a vector potential $C_\dm$ and a spatial metric $k^{\m\n}$, which in appropriate coordinates correspond to a relativistic metric 
\begin{eqnarray*}
ds^2=g_{\m\n}dx^\m dx^\n&=&-e^{-\Psi}(c\, dt+C_i dx^i)^2+e^\Psi k_{ij}dx^idx^j\\&&
+\calo(c)\, dt^2+\calo_i(c^{0})dtdx^i+\calo_{ij}(c^{-1})dx^idx^j
\end{eqnarray*}
In case the fields $(\Psi, C_\dm, k^{\m\n})$ are time independent the leading part of the metric above describes a generic stationary metric and indeed the leading order nonrelativistic equations in table \ref{eqtab} can be identified with the Einstein equations for stationary metrics. Absence of time derivatives in the leading order equations in table \ref{eqtab} implies that more generally at leading order in the large $c$ expansion any solution to Einstein equations will be of the form of a stationary metric, but with integration constants replaced by an arbitrary function of time. This extra time dependence will lead to source terms in the subleading equations. This observation gives the large $c$ expansion the interpretation of an expansion around the stationary sector of GR. This generalizes the observation of \cite{VandenBleeken:2019gqa} that the large $c^2$-expansion has the interpretation of an expansion around the static sector of GR. 

If one chooses the leading order stationary metric to be Minkowski space the large $c$ expansion reduces to the standard post-Newtonian one. But more generally one can choose any stationary metric (with time-dependent integration constants) as the starting point for the expansion, extending it into the strong gravitational regime. We illustrate this explicitly using the example of the Kerr metric.

In our derivation of the leading order equations we make use of the Newton-Cartan formalism. We make a number of comments and observations that are new with respect to \cite{Dautcourt:1996pm, Tichy:2011te, VandenBleeken:2017rij, Hansen:2020}. One main novelty is that we leave the leading time-like component of the relativistic metric -- denoted $\os{A}{-2}$ -- free, rather than choosing it to be $-1$ as was done previously. This clarifies the structure of the various potentials and equations appearing at arbitrary order, see table \ref{Atab}. Additionaly there now appears an extra local scaling symmetry under which $\os{A}{-2}=-1$ is a particular choice of gauge. We point out that there is another appealing choice of gauge, namely one where the torsion of the Newton-Cartan structure vanishes.

In addition to this new scaling symmetry we also consider the Milne boost symmetry inherent to the Newton-Cartan formalism and in particular carefully analyze the diffeomorphism symmetry at various orders. This leads us to introduce a new set of variables and to reorganize the equations, see table \ref{vartab}, simplifying them greatly. This suggests that, especially if one would attempt to go to higher orders, the large $c$ expansion can be formulated more efficiently in an alternative set of variables rather then those used in \cite{Dautcourt:1996pm, Tichy:2011te, VandenBleeken:2017rij, Hansen:2020}. 

The paper is organized as follows. In section \ref{prepsec} we review some of the Newton-Cartan formalism we need, introduce a convenient dotted index notation and prepare GR towards a covariant nonrelativistic expansion. Section \ref{expsec} concerns itself with this expansion: we define it precisely, discuss its symmetries and present the leading order equations upon the inclusion of odd terms, our main result. We then comment on various properties of these equations in section \ref{disc}. We present an example, discuss various gauge choices and provide a variational principle. Appendices \ref{TTapp} and \ref{A2B} contain some technicalities, while appendix \ref{algapp} contains a three parameter generalization of the nonrelativistic algebra of \cite{Hansen:2018ofj, Hansen:2020}, which we believe is new. We take the non-uniqueness of the nonrelativistic local translation algebra as an indication that the large $c$ expansion is most naturally expressed in terms of metric variables transforming under expanded diffeomorphism symmetries rather than vielbein variables transforming under expanded Poincar\'e symmetries.

\section{GR in Newton-Cartan split form}\label{prepsec}
A diffeomorphism invariant formulation of the nonrelativistic approximation to GR -- the large $c$ expansion -- is naturally formulated using the language of Newton-Cartan geometry \cite{Dautcourt:1996pm, Tichy:2011te, VandenBleeken:2017rij,Hansen:2020}. One feature of Newton-Cartan geometry is that it provides a local split of time versus space, something expected in a nonrelativistic theory. 

Anticipating these facts we choose to 'prepare' GR towards this expansion by performing this split in space vs time already at the relativistic level. We do this by introducing an arbitrary Newton-Cartan structure $(\tau_\m, h^{\m\n})$ in addition to the relativistic metric $g_{\m\n}$. Later, once we perform the expansion, this artificial Newton-Cartan structure will be given a natural interpretation by equating it to some of the leading order components of $g_{\m\n}$. Our approach is similar in spirit to that of \cite{Hansen:2019svu, Hansen:2020}, but differs in that the Newton-Cartan structure we introduce will be chosen to be $c$ independent.

In this section we first discuss how any tensorial quantity can be split along temporal and spatial directions using a Newton-Cartan structure and introduce a dotted index notation to represent this split in a convenient way. In the second part of the section we apply such a split to the main ingredients of GR: the metric, energy-momentum tensor and their equations of motion.  

\subsection{Newton-Cartan split: generalities}\label{decompsec}
We will restrict the discussion in this paper to 1+3 dimensions -- a generalization to arbitrary dimensions is straightforward -- and work with coordinates $x^\mu$, $\mu=0,1,2,3$. 

We'll define a Newton-Cartan structure $(\tau_\m, h^{\m\n})$ as a positive semi-definite, symmetric 2-tensor $h^{\m\n}$ together with a zero eigencovector $\tau_\mu$ that is unique up to rescaling, i.e.
\begin{equation}
h^{\m\n}\omega_\nu=0\quad \Leftrightarrow\quad \omega_\mu\sim\,\tau_\mu\,.
\end{equation}
The one-form $\tau_\mu$ is often referred to as the clock form as it sets the direction of time, while $h^{\m\n}$ encodes a purely spatial metric \cite{Cartan1,Cartan2}.

Given a Newton-Cartan structure one can introduce additional fields $\tau^\m$ and $h_{\m\n}$ such that
\begin{equation}
\tau_\mu\tau^\nu+ h_{\m\r}h^{\r\n}=\delta_\m^\n \qquad \tau^\m \tau^\n h_{\m\n}=0\,. \label{splitform}
\end{equation}
Note that $\tau^\m$ and $h_{\m\n}$ are not unique for a given Newton-Cartan structure. This non-uniqueness can be interpreted as a gauge invariance of the formalism, the Milne boosts that we will discuss below. The equations \eqref{splitform} can also be read as the definition of two complementary projectors:
\begin{equation}
\tau_\m^\n=\tau_\m\tau^\n\qquad h_\m^\n= h_{\m\r}h^{\r\n} 
\end{equation}
Using these, any tensor index can be decomposed into a temporal and spatial part, something we will refer to as a 'Newton-Cartan split'. For an arbitrary one-form $U_\mu$ for example, such decomposition reads 
\begin{equation}
U_\mu=\tau_\m\tau^\n U_\nu+ h_\m^\n U_\nu\label{decomp1}
\end{equation}
Since we will perform such a Newton-Cartan split on essentially any tensor we will encounter it will be useful to introduce some more compact notation. We'll write the temporal part of $U_\mu$ as 
\begin{equation}
U = \tau^\m U_\m
\end{equation}
and we will indicate the spatial projection by putting a dot on the projected index:
\begin{equation}
 U_\dm = h^\n_\m U_\n
\end{equation}
In this notation the decomposition \eqref{decomp1} then simplifies to
\begin{equation}
U_\m = \tau_\m U + U_\dm
\end{equation}
Note that dotted indices contracted with $\tau$ vanish, while the spatial components can be lowered and raised by $h$
\begin{equation}
\tau_\m U^\dm = 0 \qquad \tau^\m U_\dm = 0 \qquad h_{\m\n}U^\dn = U_\dm \qquad h^{\m\n}U_\dn = U^\dm\label{indexrels}
\end{equation}
This notation can be safely extended to higher rank tensors as long as they are either fully symmetric or anti-symmetric. For example for a (1,2)-tensor that is symmetric in its upper indices we get
\begin{equation}
V_\m{}^{\n\r}=V\tau_\m\tau^\n\tau^\r+2V^{(\dn}\tau^{\r)}_\m+V^{\dn\dr}\tau_\mu +V_{\dm}\tau^\n\tau^\r+2V_{\dm}{}^{(\dn}\tau^{\r)}+V_\dm{}^{\dn\dr}
\end{equation}
We will reserve the raising and lowering of dotted indices with $h$ as in \eqref{indexrels} only for tensors where there will be no ambiguity in the notation. For example, note that in the split of $V_\m{}^{\n\r}$ above we have $V^{\dm\dn}=\tau^\r h^\m_\s h^\n_\l V_\r{}^{\s\l}$ and $V_{\dm}{}^{\dn}=\tau_\r h_{\m}^\s h^\n_\l V_\s{}^{\r\l}$ and thus $V^{\dm\dn}\neq h^{\m\r}V_{\dr}{}^{\dn}$.   

Finally it is important to point out that we will not work with the most general type of Newton-Cartan structure, but rather assume that the clock form satisfies $\partial_{[\dm}\tau_{\dn]}=0$. This condition can 
be expressed in three equivalent ways:
 \begin{equation}
 \tau_{[\mu}\partial_\n\tau_{\r]}=0\quad\Leftrightarrow\quad \partial_{[\dm}\tau_{\dn]}=0\quad\Leftrightarrow\quad \partial_{[\m}\tau_{\n]}=\tau_{[\mu}a_{\nu]}\,,\quad a_\mu=a_\dm\label{tlcond}
 \end{equation}
Geometrically these conditions guarantee the existence of a foliation by spatial hypersurfaces. Another interpretation is that this condition restricts the torsion of any connection compatible with the Newton-Cartan structure and for this reason \eqref{tlcond} also goes under the name of twistless torsion \cite{Christensen:2013lma}. We make this assumption on the Newton-Cartan structure from the beginning, anticipating compatibility with the expanded Einstein equations \cite{VandenBleeken:2017rij}. 

Note that from \eqref{tlcond} it follows that
\begin{equation}
a_\dm=2\tau^\r\partial_{[\r}\tau_{\m]}=L_{\tau}\tau_\m\label{adef}
\end{equation}
where furthermore
\begin{equation}
\partial_{[\dm}a_{\dn]}=0\label{spatialclosed}
\end{equation}
 so that locally
 \begin{equation}
 a_\dm=\partial_{\dm}\psi\label{apart}
 \end{equation}
The derivation of \eqref{tlcond} and the other formulae above is shortly reviewed in appendix \ref{TTapp}.

\subsection{Newton-Cartan split: GR}
We take as our dynamical variables for GR a Lorentzian metric $g_{\m\n}$ and the trace-reversed energy-momentum tensor $\calt_{\m\n}$:
\begin{equation}
\calt_{\m\n}=c^{-4}\left(T_{\m\n}-\frac{1}{2}g_{\m\n}T_\r{}^\r\right)
\end{equation}
The dynamics is then provided by the Einstein equations, which we write as
\begin{equation}
\cale_{\m\n}=R_{\m\n}-8\pi \GN\, \calt_{\m\n}=0\label{Ees}
\end{equation}
The Bianchi identity satisfied by the Ricci tensor $R_{\m\n}$ guarantees the conservation of energy-momentum, which is equivalent to
\begin{equation}
\calc_{\m}=\nabla^\r \calt_{\r\m}-\frac{1}{2}\partial_\m\calt_\r{}^{\r}=0\label{emc}
\end{equation}
We then apply a Newton-Cartan split to each of these ingredients:
\begin{eqnarray}
\calt_{\m\n}&=&\calt \tau_{\m}\tau_\n+2\tau_{(\m}\calt_{\dn)}+\calt_{\dm\dn}\\
\cale_{\m\n}&=&\cale \tau_{\m}\tau_\n+2\tau_{(\m}\cale_{\dn)}+\cale_{\dm\dn}\\
\calc_\m&=&\calc \tau_\m+\calc_\dm
\end{eqnarray}
We will use a separate notation for the split of the metric and its inverse:
\begin{eqnarray} \label{metricdecompose}
g_{\m\n}&=&A\tau_\mu\tau_\nu+A_\dm\tau_\n+A_\dn\tau_\mu+B_{\dm\dn} \\
g^{\m\n}&=&B\tau^\m\tau^\n+B^\dm\tau^\n+B^{\dn} \tau^\mu+A^{\dm\dn}\nonumber
\end{eqnarray}
This $A$ vs $B$ notation is introduced to indicate that we will treat the $A$ variables as independent fields, while the $B$ fields are interpreted as fully determined in terms of the $A$'s through the condition that $g^{\m\r}g_{\r\n}=\delta_\n^\m$\,:
\begin{eqnarray}
B&=&A^{-1}(1+A^{-1}A_\dm A_\dn A^{\dm\dn})\nonumber\\
B^\dm&=&-A^{-1} A_\dn A^{\dn\dm}\label{Ba}\\
B_{\dm\dn}&=&A_{\dm\dn}+A^{-1}A_\dm A_\dn\nonumber
\end{eqnarray}
with $A_{\dm\dn}$ is the unique\footnote{Contrary to \eqref{splitform}, which has no unique solution for $\tau^\mu$ and $h_{\m\n}$ given $\tau_\mu$ and $h^{\m\n}$, the equation \eqref{Aeq} has a unique solution for $A_{\dm\dn}$ given $\tau_\mu\,, \tau^\mu$ and $A^{\dm\dn}$.} solution to
\begin{equation}
\tau_\m\tau^\n+A_{\dm\dr}A^{\dr\dn}=\delta_\m^\n\label{Aeq}
\end{equation}
The upshot of this Newton-Cartan split of GR can be found in table \ref{splittab}.

\begin{table}
	\framebox[\linewidth]{\begin{minipage}{0.9\linewidth}
			\begin{center}
				{\bf Dynamical variables}
			\end{center}
			\begin{equation*}
			A\,,\quad A_\dm\,,\quad A^{\dm\dn}\qquad\mbox{and}\qquad \calt\,,\quad\calt_\dm\,,\quad \calt_{\dm\dn}\,.
			\end{equation*}
			\begin{center}
				{\bf Equations}
			\end{center}
			\begin{equation*}
			\cale=0\,,\quad \cale_\dm=0\,,\quad \cale_{\dm\dn}=0\qquad\mbox{and}\qquad \calc=0\,,\quad\calc_\dm=0
			\end{equation*}
			\caption{\label{splittab} GR formulated through a Newton-Cartan split.}
			
			\vspace{0.1cm}
			
		\end{minipage}}
	\end{table}

\section{The large $c$ expansion}\label{expsec}
A manifestly diffeomorphism invariant approximation to GR can be constructed by expanding the relativistic metric and its inverse in inverse powers of the speed of light \cite{Dautcourt:1996pm,Tichy:2011te,VandenBleeken:2017rij,Hansen:2020}:
\begin{equation}
g_{\m\n}(c) = \sum_{k=-2}^{\infty} \os{g}{k}_{\m\n} c^{-k} \qquad g^{\m\n}(c) = \sum_{k=0}^{\infty} \os{g}{k}^{\m\n} c^{-k}\label{metexp}
\end{equation}
Although it is consistent to assume all coefficients of odd powers to vanish -- as was done in \cite{Dautcourt:1996pm,Tichy:2011te,VandenBleeken:2017rij,Hansen:2020} -- we will explore in this work the consequences of relaxing this assumption. The presence of non-vanishing coefficients for odd powers of $c$ is motivated by the fact that these coefficients can be sourced in the equations of motion by certain types of energy-momentum \cite{VandenBleekenx, Hansen:2020}.

We will first discuss some generalities and then focus on the leading order, working out explicitly the dynamical equations to this order. Symmetries will play an important role in an appropriate organization of the result.

\subsection{Setup and general observations}\label{setsec}
We will take a slightly different -- but equivalent -- approach than the one that was previously taken in \cite{Dautcourt:1996pm,Tichy:2011te,VandenBleeken:2017rij,Hansen:2020}. We will perform the expansion not directly in terms of the relativistic metric as in \eqref{metexp}, but rather expand the components as obtained after a Newton-Cartan split as in \eqref{metricdecompose}, see also table \ref{splittab}. It should be stressed that the Newton-Cartan structure $\tau_\mu\,, h^{\m\n}$ is taken to be independent of $c$.

Our expansion ansatz for the dynamical fields is then
\begin{align}
A(c) &= \sum^{\infty}_{k=-2} \os{A}{k} c^{-k} &\qquad A_\dm(c) &= \sum^{\infty}_{k=-1} \os{A}{k}_\dm c^{-k} & \qquad A^{\dm\dn}(c) & = \sum^{\infty}_{k=0} \os{A}{k}^{\dm\dn} c^{-k}\\
\calt(c) &= \sum^{\infty}_{k=-2} \os{\calt}{k} c^{-k} &\qquad \calt_\dm(c) &= \sum^{\infty}_{k=-1} \os{\calt}{k}_\dm c^{-k} & \qquad \calt_{\dm\dn}(c) & = \sum^{\infty}_{k=0} \os{\calt}{k}_{\dm\dn} c^{-k} \label{expans}
\end{align}
This ansatz is based on a number of starting assumptions, equivalent to those of \cite{Dautcourt:1996pm,Tichy:2011te,VandenBleeken:2017rij,Hansen:2020} with the exception that odd powers are allowed to be non-vanishing. In those previous works \cite{Dautcourt:1996pm,Tichy:2011te,VandenBleeken:2017rij,Hansen:2020} the choice $\os{A}{-2}=-1$ was made, but as we will explain below that is simply a choice of gauge for a local scaling symmetry in our formulation. Leaving $\os{A}{-2}$ free has some advantages. In particular it makes manifest the fact that at each order in the expansion there appear two new triplets of fields, ($\os{A}{k}\,, \os{A}{k+1}_\dm\,, \os{A}{k+2}^{\dm\dn}$) and ($\os{\calt}{k}\,, \os{\calt}{k+1}_\dm\,, \os{\calt}{k+2}_{\dm\dn}$), all of which are a priori free to vary. In table \ref{Atab} we compare the potentials as we define them here, to the potentials as previously considered in \cite{Dautcourt:1996pm,Tichy:2011te,VandenBleeken:2017rij,Hansen:2020}. In the formulation using the Newton-Cartan split it also becomes clear that the proper organization of the orders is not just by counting inverse powers of $c$, but rather that at a given order the powers of $c$ depend on the number of spatial indices the field carries. The triplet ($\os{A}{k}\,, \os{A}{k+1}_\dm\,, \os{A}{k+2}^{\dm\dn}$) for example consists of coefficients of the powers $(c^{-k},c^{-k-1},c^{-k-2})$. This is something which was not properly appreciated in \cite{Dautcourt:1996pm,Tichy:2011te,VandenBleeken:2017rij}, but suggested by the results of \cite{Hansen:2018ofj,Hansen:2020}. That this is indeed the more appropriate point of view is supported by the fact that the equations of motion -- which we will further work out below -- organize themselves in the following form\footnote{Note that the structure of \eqref{eqhier} follows upon assuming that $\os{R}{-4}_{\m\n}=0$ which amounts to the twistless torsion condition \eqref{tlcond}, see \cite{VandenBleeken:2017rij}. But since this is equivalent to the assumption that $\os{\calt}{-4}_{\m\n}=\os{\calt}{-3}_{\m\n}=0$, one sees that this condition is already encoded in \eqref{expans}.\label{twistfoot}}:
\begin{eqnarray}
\os{\cale}{k}[\os{A}{\leq k}\,, \os{A}{\leq k+1}_\dm\,, \os{A}{\leq k+2}^{\dm\dn};\os{\calt}{k}]&=&0\nonumber\\
\os{\cale}{k+1}_\dm [\os{A}{\leq k}\,, \os{A}{\leq k+1}_\dm\,, \os{A}{\leq k+2}^{\dm\dn};\os{\calt}{k+1}_\dm]&=&0\label{eqhier}\\
\os{\cale}{k+2}_{\dm\dn} [\os{A}{\leq k}\,, \os{A}{\leq k+1}_\dm\,, \os{A}{\leq k+2}^{\dm\dn};\os{\calt}{k+2}_{\dm\dn}]&=&0\nonumber
\end{eqnarray}
One sees that if the expansion is truncated at order $n$, i.e. keeping only triplets of the form ($\os{A}{k}\,, \os{A}{k+1}_\dm\,, \os{A}{k+2}^{\dm\dn}$) and ($\os{\calt}{k}\,, \os{\calt}{k+1}_\dm\,, \os{\calt}{k+2}_{\dm\dn}$) with $k\leq n$, then the above set of equations for $k\leq n$ is consistent. Furthermore the set of equations has a hierarchic structure, in that one can solve them recursively in the order. The leading order corresponds to $n=-2$ and will be worked out fully below.

Finally we should point out that since $\tau_\m \os{A}{0}^{\dm\dn}=0$ we can choose to identify the spatial part of the Newton-Cartan structure introduced in section \eqref{decompsec} with it:
\begin{equation}
\os{A}{0}^{\dm\dn}=h^{\m\n}\label{hdef}
\end{equation}
Translated back to the more familiar relativistic metric variables this amounts to
\begin{equation}
g_{\m\n}=c^2 \os{A}{-2}\tau_\m\tau_\n+\calo(c)\qquad g^{\m\n}=h^{\m\n}+\calo(c^{-1})
\end{equation}
So once we start considering the large $c$ expansion, we identify the Newton-Cartan structure that was artificially introduced in section \ref{decompsec} with the leading coefficients of the relativistic metric and its inverse.

\begin{table}[!ht]
	\begin{mdframed}
	\begin{center}
	\begin{tabular}{l|ccc}
		LO & $\os{A}{-2}=-1$& $\os{A}{-1}_{\dm}=0$ & \textcolor{violet}{$\os{A}{0}^{\dm\dn}=h^{\m\n}$}\\
		N$^{1/2}$LO&  $\os{A}{-1}=0$& \textcolor{violet}{$\os{A}{0}_\dm=m_\dm$}& $\os{A}{1}^{\dm\dn}=0$\\
		NLO (Newtonian)& \textcolor{blue}{$\os{A}{0}=-2\Phi$}& $\os{A}{1}_\dm=0$ & \textcolor{red}{$\os{A}{2}^{\dm\dn}=\Phi^{\dm\dn}$}\\
		N$^{3/2}$LO&$\os{A}{1}=0$& $\textcolor{red}{\os{A}{2}_\dm=\gamma_\dm}$ & $\os{A}{3}^{\dm\dn}=0$\\
		N$^2$LO&$\os{A}{2}=\textcolor{red}{-\gamma}$& $\os{A}{3}_\dm=0$ & $\os{A}{4}^{\dm\dn}=0$\\
	\end{tabular}\caption{In this table all gravitational potentials appearing up to next to next to leading order (N$^2$LO) in the large $c$ expansion are listed. In the previous literature on this expansion \cite{Dautcourt:1996pm,Tichy:2011te,VandenBleeken:2017rij,Hansen:2020} most of these potentials have been assumed to vanish, as indicated in the table. The NLO level can be identified as the Newtonian level, since $\Phi$ (in blue) is Newton's potential. Up to this order we have used the nomenclature of \cite{Hansen:2020} for the fields. For the two highest orders we used the nomenclature of \cite{Tichy:2011te}. Non-trivial values for the fields in violet have only been considered more recently \cite{VandenBleeken:2017rij,Hansen:2020}. If the potentials in violet are assumed to vanish then the potentials in red can be identified with those appearing at first Post-Newtonian (PN) order in the standard PN expansion \cite{Tichy:2011te} . In this paper we will focus on the LO only but will allow non-trivial values for the full triplet, see table \ref{ourA}.
} \label{Atab}
\end{center}
\end{mdframed}
	
\end{table}

\subsection{Symmetries}
The relativistic theory we start from, GR, has as its symmetries simply the diffeomorphisms\footnote{In a frame formulation one can introduce local Poincare symmetries. These symmetries are however not compatible with the large $c$ expansion ansatz. This is related to the appearance of exotic nonrelativistic algebras that have no relativistic origin, see appendix \ref{algapp}. For this reason we refrain from using a frame formulation in this work.}. These diffeomorphisms are however allowed to depend on the speed of light $c$, which implies that they will generate a new independent set of symmetries at each order of the expansion \cite{Tichy:2011te, VandenBleeken:2017rij}. In addition to the diffeomorphisms -- and their descendants -- there are two additional symmetries that appear into the expansion simply because the Newton-Cartan structure and decomposition we introduced in section \ref{prepsec} have some redundancy: a nonrelativistic local boost symmetry and a local scaling symmetry. We now discuss the various symmetries in turn, in the next subsection they will be used to organize the LO equations.

\subsubsection*{Milne Boosts}
Given a Newton-Cartan structure $(\tau_\m,h^{\m\n})$ the 'inverse' fields $\tau^\mu$ and $h_{\m\n}$ -- defined to satisfy \eqref{splitform} -- are not unique. An equivalent set of solutions to \eqref{splitform} is generated through the infinitesimal transformations
\begin{equation}
\d_\chi \tau^\m = -\chi^{\dm} \qquad \d_\chi h_{\m\n} = \chi_\dm \tau_\n + \chi_\dn \tau_\m\label{NCboost}
\end{equation}
Note that this implies\footnote{One computes $\delta_\chi (\partial_\dm \psi)=\tau_\m \chi^\dr \partial_\dr \psi+\partial_\dm \delta_\chi\psi$.} via (\ref{adef}, \ref{apart}) that
\begin{equation}
\delta_\chi a_\dm=\tau_\m \chi^\dr a_\dr\qquad \delta_\chi \psi=0 \label{psiboost}
\end{equation}

Since the relativistic metric $g_{\m\n}$ is independent of our choice of $\tau^\m$ and $h_{\m\n}$ used to split it, it follows that the components as defined in \eqref{metricdecompose} must transform as 
\begin{equation}
\d_{\chi}A = -2\chi^\dm A_\dm  \qquad \d_{\chi}A_\dm = (\tau_\m A_\dr -B_{\dm\dr})\chi^\dr\qquad  \d_{\chi} A^{\dm\dn}=2B^{(\dm} \chi^{\dn)}
\end{equation}
Since we choose $\tau_\m\,, \tau^\m\,, h^{\m\n}$ and $h_{\m\n}$ to be $c$ independent it follows that the parameter $\chi_\dm$ will be $c$ independent and thus the action of the boost symmetry on the expanded fields is immediate:
\begin{equation}
\d_{\chi}\os{A}{k} =-2\chi^\dm  \os{A}{k}_\dm \qquad \d_{\chi}\os{A}{k}_\dm = (\tau_\m \os{A}{k}_\dr -\os{B}{k}_{\dm\dr})\chi^\dr \qquad  \d_{\chi} \os{A}{k}^{\dm\dn}=2\os{B}{k}^{(\dm} \chi^{\dn)}
\end{equation}
Still, these are rather complicated transformations due to the rather lengthy expressions for the $B$'s in terms of the $A$'s once the order increases, see appendix \ref{A2B}.  On the leading triplet the action is very simple however:
\begin{equation}
\delta_\chi \os{A}{-2}=0\qquad \delta_\chi \os{A}{-1}_\dm=\tau_\mu \os{A}{-1}_\dr\chi^\dr\qquad  \delta_\chi \os{A}{0}^{\dm\dn}=\delta_\chi h^{\m\n}=0 \label{tripboost}
\end{equation}
The boost transformations of the leading order energy-momentum triplet are found to be
\begin{equation}
\delta_\chi \os{\calt}{-2}=0\qquad \delta_\chi \os{\calt}{-1}_\dm=\tau_\mu \os{\calt}{-1}_\dr\chi^\dr\qquad  \delta_\chi \os{\calt}{0}_{\dm\dn}=(\tau_{\mu}\os{\calt}{0}_{\dn\dr}+\tau_{\nu}\os{\calt}{0}_{\dm\dr})\chi^\dr \label{Ttripboost}
\end{equation}

\subsubsection*{Scaling}
A re-scaling of the clock form can be absorbed in the definitions of the $A$ fields, while leaving $g_{\m\n}$ invariant. Infinitesimally these scaling transformations that leave \eqref{metricdecompose} and \eqref{splitform} invariant are
\begin{equation}
\delta_\lambda \tau_\mu=\lambda \tau_\mu\,,\quad \delta_\lambda \tau^\mu=-\lambda \tau^\mu\,,\label{NCscal}
\end{equation}
\begin{equation*}
	\delta_\lambda A=-2\lambda A\,,\quad \delta_\lambda A_\dm=-\lambda A_\dm\,,\quad \delta_\lambda A^{\dm\dn}=0
\end{equation*}
Let us point out that this implies that 
\begin{equation}
\delta_\lambda a_\dm=-\partial_\dm\lambda \label{ascal}
\end{equation}
Because of \eqref{apart} this reveals that the twistless torsion degree of freedom in $\tau_\mu$ is pure gauge, something which we'll discuss further below.  

Due to the $c$ independence of $\lambda$ it follows that the scaling acts straightforwardly on the expansion coefficients:
\begin{equation}
\delta_\lambda \os{A}{k}=-2\lambda \os{A}{k}\,,\quad \delta_\lambda \os{A}{k}_\dm=-\lambda\label{tripscal} \os{A}{k}_\dm\,,\quad \delta_\lambda \os{A}{k}^{\dm\dn}=0\,.
\end{equation}
Similarly
\begin{equation}
\delta_\lambda \os{\calt}{k}=-2\lambda \os{\calt}{k}\,,\quad \delta_\lambda \os{\calt}{k}_\dm=-\lambda\label{Ttripscal} \os{\calt}{k}_\dm\,,\quad \delta_\lambda \os{\calt}{k}_{\dm\dn}=0\,.
\end{equation}

\subsubsection*{Diffeomorphisms}
One of the key features of GR is its invariance under diffeomorphisms. A priori they can  depend arbitrarily on the speed of light $c$. Compatibility with the expansion ansatz \eqref{metexp} requires however that the generating vector field satisfies \cite{Tichy:2011te, VandenBleeken:2017rij}
\begin{equation}
\xi^\mu(c)=\sum_{k=0}^\infty \os{\xi}{k}^\mu c^{-k}
\end{equation}
The zeroth order coefficients $\os{\xi}{0}^\mu$ generate the diffeomorphisms of the nonrelativistic theory obtained by the expansion, while the higher order coefficients generate additional gauge transformations. The action on the coefficients of an arbitrary relativistic tensor $U_{\m_1\ldots\m_n}^{\n_1\ldots\n_m}(c)$ is
\begin{equation}
\delta_{\os{\xi}{k}} \os{U}{l}_{\m_1\ldots\m_n}^{\n_1\ldots\n_m}=L_{\os{\xi}{k}}\os{U}{l-k}_{\m_1\ldots\m_n}^{\n_1\ldots\n_m}
\end{equation}

We define the Newton-Cartan structure $\tau_\m$, $h^{\m\n}$ and $\tau^\m$, $h_{\m\n}$ to transform as tensors under $\os{\xi}{0}^\mu$ but to be invariant under all subleading diffeomorphisms $\os{\xi}{k}^\mu$, $k>0$. We stress that although such definition is consistent, it implies that the components $A$, $A_{\dm}$ and $A^{\dm\dn}$ do {\it not} transform as tensors under $c$ dependent diffeomorphisms. Rather their transformations are defined as the respective components of the transformed relativistic metric:
\begin{equation}
\delta_{\os{\xi}{k}}\os{g}{l}_{\m\n}=\tau_\m\tau_\n\delta_{\os{\xi}{k}}\os{A}{l}+2\tau_{(\m}\delta_{\os{\xi}{k}}\os{A}{l}_{\dm)}+\delta_{\os{\xi}{k}}\os{B}{l}_{\dm\dn}\qquad \quad (k>0)
\end{equation}
This is equivalent to
\begin{eqnarray}
\delta_{\os{\xi}{k}}\os{A}{l}&=&\tau^\r\tau^\s \delta_{\os{\xi}{k}}\os{g}{l}_{\r\s}=\tau^\r\tau^\s L_{\os{\xi}{k}}\os{g}{l-k}_{\r\s}\nonumber\\
\delta_{\os{\xi}{k}}\os{A}{l}_\dm&=&\tau^\r h^\s_\m \delta_{\os{\xi}{k}}\os{g}{l}_{\r\s}=\tau^\r h^\s_\m L_{\os{\xi}{k}}\os{g}{l-k}_{\r\s}\\
\delta_{\os{\xi}{k}}\os{A}{l}^{\dm\dn}&=&h_\r^\m h_\s^\n \delta_{\os{\xi}{k}}\os{g}{l}^{\r\s}=h_\r^\m h_\s^\n L_{\os{\xi}{k}}\os{g}{l-k}^{\r\s}\nonumber
\end{eqnarray}
which in turn can be rewritten via \eqref{metricdecompose} as
\begin{eqnarray}
\delta_{\os{\xi}{k}}\os{A}{l}&=&L_{\os{\xi}{k}}\os{A}{l-k}+2\os{A}{l-k}(\partial\os{\xi}{k}-a_\dr \os{\xi}{k}^\dr )-2 \os{A}{l-k}_\dr L_{\os{\xi}{k}}\tau^\r \qquad\qquad (k>0)\nonumber\\
\delta_{\os{\xi}{k}}\os{A}{l}_\dm&=&\os{A}{l-k}(\partial_\dm\os{\xi}{k}+a_\dm \os{\xi}{k})+h^\s_\dm L_{\os{\xi}{k}}\os{A}{l-k}_\ds+\os{A}{l-k}_\dm(\partial\os{\xi}{k}-a_\dr \os{\xi}{k}^\dr )-\os{B}{l-k}_{\dm\dr}L_{\os{\xi}{k}}\tau^\r\nonumber\\
\delta_{\os{\xi}{k}}\os{A}{l}^{\dm\dn}&=&2\os{B}{l-k}^{(\dm}h^{\dn)}_\r L_{\os{\xi}{k}}\tau^\r +h_\r^\m h_\s^\n L_{\os{\xi}{k}}\os{A}{l-k}^{\dr\ds}\nonumber
\end{eqnarray}
Note that because $\os{A}{k}=\os{A}{k+1}_\dn=\os{A}{k+2}^{\dm\dn}=0$ when $k<-2$ it follows that the subleading diffeomorphism $\os{\xi}{k}$ acts non-trivially only on the triplets $(\os{A}{l},\os{A}{l+1}_\dn,\os{A}{l+2}^{\dm\dn})$ for which $l\geq k-2$. In particular, at leading order only $\os{\xi}{1}$ acts non-trivially, in the simple fashion
\begin{equation}
\delta_\os{\xi}{1}\os{A}{-2}=0\qquad \delta_\os{\xi}{1}\os{A}{-1}_\dm=\os{A}{-2}(\partial_\dm\os{\xi}{1}+a_\dm \os{\xi}{1})\qquad \delta_{\os{\xi}{1}}\os{A}{0}^{\dm\dn}=0 \label{tripsdif}
\end{equation}
Note that the transformation of $\os{A}{-1}_\dm$ resembles a U(1) transformation, and indeed we will recast it as such below. Furthermore, the spatial part $\os{\xi}{1}^\dm$ acts trivially at this order. Let us point out that this simple structure repeats itself at all orders, when one considers only the action of the highest order diffeomorphisms on the highest order triplet:
\begin{equation}
\delta_\os{\xi}{l+3}\os{A}{l}=0\qquad \delta_\os{\xi}{l+3}\os{A}{l+1}_\dm=\os{A}{-2}(\partial_\dm\os{\xi}{l+3}+a_\dm \os{\xi}{l+3})\qquad \delta_{\os{\xi}{l+3}}\os{A}{l+2}^{\dm\dn}=0
\end{equation}

A similar analysis reveals the transformations of the leading order energy-momentum triplet to be
\begin{equation}
\delta_\os{\xi}{1}\os{\calt}{-2}=0\qquad \delta_\os{\xi}{1}\os{\calt}{-1}_\dm=\os{\calt}{-2}(\partial_\dm\os{\xi}{1}+a_\dm \os{\xi}{1})\qquad \delta_{\os{\xi}{1}}\os{\calt}{0}_{\dm\dn}=2\os{\calt}{-1}_{(\dm}\partial_{\dn)}\os{\xi}{1}+2\os{\calt}{-1}_{(\dm}a_{\dn)} \os{\xi}{1}\label{Ttripsdif}
\end{equation}

\subsection{Leading Order}
At leading order in the large $c$ expansion we have the dynamical fields ($\os{A}{-2}\,, \os{A}{-1}_\dm\,, \os{A}{0}^{\dm\dn}$) and ($\os{\calt}{-2}\,, \os{\calt}{-1}_\dm\,, \os{\calt}{0}_{\dm\dn}$). The leading order in the expansion of the equations in table \ref{splittab} provides a consistent set of equations for these fields that we will now compute. Although similar in spirit to \cite{Dautcourt:1996pm, Tichy:2011te, VandenBleeken:2017rij, Hansen:2020}, this is the first such calculation keeping $\os{A}{-1}_{\dm}$ arbitrary.

\subsubsection*{Field redefinitions}
A brute force calculation -- using \eqref{tlcondapp}, see footnote \ref{twistfoot} -- reveals that the leading coefficients of the Einstein and conservation equations are
\begin{equation}
\os{\cale}{-2}=0\,,\quad \os{\cale}{-1}_\dm=0\,,\quad \os{\cale}{0}_{\dm\dn}=0\,,\qquad \os{\calc}{-1}=0\qquad \os{\calc}{0}_\dm=0 \label{LOeq}
\end{equation}
Although these equations can in principle be expressed as equations for the variables ($\os{A}{-2}\,, \os{A}{-1}_\dm\,, \os{A}{0}^{\dm\dn}$) and ($\os{\calt}{-2}\,, \os{\calt}{-1}_\dm\,, \os{\calt}{0}_{\dm\dn}$) they can be much more efficiently formulated in terms of an equivalent set of equations expressed in terms of redefined variables that are more adapted to the symmetries. Schematically we replace
\begin{align}
\quad \qquad \os{A}{-2}&\rightarrow \Psi &\quad \os{A}{-1}_\dm &\rightarrow C_\dm & \quad \os{A}{0}^{\dm\dn}&=h^{\m\n} \rightarrow k^{\m\n}\\
\quad \qquad \os{\calt}{-2}&\rightarrow \mathfrak{T}&\quad  \os{\calt}{-1}_\dm &\rightarrow \mathfrak{T}_\dm & \quad \os{\calt}{0}_{\dm\dn}&\rightarrow \mathfrak{T}_{\dm\dn}\label{fdef}
\end{align}
and 
\begin{gather}
\os{\cale}{-2} \rightarrow \mathfrak{E} \qquad \qquad \os{\cale}{-1}_\dm \rightarrow \mathfrak{E}_\dm \qquad \qquad  \os{\cale}{0}_{\dm\dn} \rightarrow \mathfrak{E}_{\dm\dn} \\
 \os{\calc}{-1} \rightarrow \mathfrak{C}\qquad \qquad \os{\calc}{0} \rightarrow \mathfrak{C}_\dm\label{edef}
\end{gather}
The detailed redefinitions behind the above schematic are collected in table \ref{vartab}. 

\begin{table}[!ht]
	\framebox[\linewidth]{\begin{minipage}{0.9\linewidth}
			\begin{eqnarray*}
				\Psi&=&2\psi-\log(-\os{A}{-2})\\
				C_\dm&=&e^{\Psi/2} (-\os{A}{-2})^{-1/2} \os{A}{-1}_\dm\\
				k^{\m\n}&=&e^{\Psi}h^{\m\n}\qquad k_{\m\n}=e^{-\Psi}h_{\m\n}\\
				&&\\
				\mathfrak{T}&=&-e^{\Psi}\os{A}{-2}^{-1}\os{\calt}{-2}\\
				\mathfrak{T}_\dm&=&e^{-\Psi/2}(-\os{A}{-2})^{-1/2}\left(\os{\calt}{-1}_\dm-\os{A}{-2}^{-1}\os{\calt}{-2}\os{A}{-1}_\dm\right)\\
				\mathfrak{T}_{\dm\dn}&=&\os{\calt}{0}_{\dm\dn}-\os{A}{-2}^{-1}(\os{A}{-1}_\dm\os{\calt}{-1}_\dn+\os{A}{-1}_\dn\os{\calt}{-1}_\dm)+\os{A}{-2}^{-2}\os{\calt}{-2}\os{A}{-1}_\dm\os{A}{-1}_\dn\\
				&&-\frac{1}{2}h_{\m\n}h^{\r\s}\left(\os{\calt}{0}_{\dr\ds}-2\os{A}{-2}^{-1}\os{A}{-1}_\dr\os{\calt}{-1}_\ds+\os{A}{-2}^{-2}\os{\calt}{-2}\os{A}{-1}_\dr\os{A}{-1}_\ds+\frac{1}{3}h_{\r\s}\os{A}{-2}^{-1}\os{\calt}{-2}\right)\\&&\\
				\mathfrak{E}&=&\os{A}{-2}^{-1}\os{\cale}{-2}\\
				\mathfrak{E}_\dm&=&(-\os{A}{-2})^{-1/2}\left(\os{\cale}{-1}_\dm-\os{A}{-2}^{-1}\os{\cale}{-2}\os{A}{-1}_\dm\right)\\
				\mathfrak{E}_{\dm\dn}&=&\os{\cale}{0}_{\dm\dn}-\os{A}{-2}^{-1}(\os{A}{-1}_\dm\os{\cale}{-1}_\dn+\os{A}{-1}_\dn\os{\cale}{-1}_\dm)+\os{A}{-2}^{-2}\os{\cale}{-2}\os{A}{-1}_\dm\os{A}{-1}_\dn\\&&
				-\frac{1}{2}h_{\m\n}h^{\r\s}\left(\os{\cale}{0}_{\dr\ds}-2\os{A}{-2}^{-1}\os{A}{-1}_\dr\os{\cale}{-1}_\ds+\os{A}{-2}^{-2}\os{\cale}{-2}\os{A}{-1}_\dr\os{A}{-1}_\ds+\frac{1}{3}h_{\r\s}\os{A}{-2}^{-1}\os{\cale}{-2}\right)
				\\
				&&\\
				\mathfrak{C}&=&(-\os{A}{-2})^{-1/2}\os{\calc}{-1}\\
				\mathfrak{C}_{\dm}&=&\os{\calc}{0}_\dm-\os{A}{-2}^{-1}\os{\calc}{-1}\os{A}{-1}_\dm
			\end{eqnarray*}
			\caption{\label{vartab} Redefinition of dynamical variables and equations.}
			
			\vspace{0.1cm}
			
	\end{minipage}}
\end{table}

We start our motivation for these redefinitions by considering the scaling symmetry. Since all variables and equations scale homogeneously with a respective weight we can make them scale invariant by  multiplication with an appropriate power $\os{A}{-2}$, which scales with weight $-2$. In particular one can check that the following objects are scale invariant: $(-\os{A}{-2})^{-1/2}\os{A}{-1}_\dm\,, \os{A}{-2}^{-1}\os{\calt}{-2}\,, (-\os{A}{-2})^{-1/2}\os{\calt}{-1}_\dm$ and  $\os{A}{-2}^{-1}\os{\cale}{-2}\,, (-\os{A}{-2})^{-1/2}\os{\cale}{-1}_\dm\,, (-\os{A}{-2})^{-1/2}\os{\calc}{-1}_\dm$.
Since after such a reformulation all equations and all variables -- except one -- are scale invariant, it follows that $\os{A}{-2}$, which scales non-trivially has to disappear. There is one subtlety to this observation, in that derivatives of $\os{A}{-2}$ will transform non-homogeneously, with the transformation including an extra $\partial_\m\lambda$. But remembering that the one-form $a_\dm$ transforms in the same way -- see \eqref{ascal} -- we can define an invariant one-form:
\begin{equation}
\Omega_{\dm}=a_\dm-\frac{1}{2}\partial_\dm \log (-\os{A}{-2})\qquad\qquad \delta_\lambda \Omega_{\dm}=0
\end{equation}
Additionally, due to \eqref{apart}, we can furthermore write
\begin{equation}
\Omega_{\dm}=\frac{1}{2}\partial_{\dm}\Psi\qquad \Psi=2\psi- \log (-\os{A}{-2})\qquad\qquad \delta_\lambda \Psi=0\label{ompart}
\end{equation}
The upshot of this observation is that the torsion one-form $a_\dm$ and the field $\os{A}{-2}$ can only appear in the scale invariant equations through the scale invariant field $\Psi$, or its derivatives.

Further simplification can be obtained by considering invariance under the subleading diffeomorphisms $\os{\xi}{1}^\mu$. The scale invariant version of $\os{A}{-1}_{\dm}$, $(-\os{A}{-2})^{-1/2}\os{A}{-1}_{\dm}$, transforms under subleading diffeomorphisms \eqref{tripsdif} as
\begin{equation}
\delta_{\os{\xi}{1}}(-\os{A}{-2})^{-1/2}\os{A}{-1}_{\dm}=-(\partial_\dm+\Omega_\dm)(-\os{A}{-2})^{1/2}\os{\xi}{1}
\end{equation}
The definition \eqref{ompart} then suggests to define
\begin{equation}
C_\dm=-e^{\Psi/2} (-\os{A}{-2})^{-1/2}\os{A}{-1}_{\dm}
\end{equation}
so that this appropriately rescaled version of $\os{A}{-1}_{\dm}$ transforms as a U(1) gauge field under subleading diffeomorphisms:
\begin{equation}
\delta_{\os{\xi}{1}}C_\dm=\partial_\dm \zeta\qquad \zeta=e^{-\Psi/2} (-\os{A}{-2})^\frac{1}{2} \os{\xi}{1}
\end{equation}
If one can make the other variables and equations manifestly invariant under these subleading diffeomorphisms then it will follow that $C_\dm$ can only appear through its gauge invariant curvature. Let us illustrate how one can indeed make the other variables invariant under $\os{\xi}{1}^\m$ transformations, by considering the transformation \eqref{Ttripsdif} of $\os{\calt}{1}_\dm$ and using the transformation \eqref{tripsdif} of $\os{A}{-1}_\dm$ one can make the invariant combination:
\begin{equation}
\delta_{\os{\xi}{1}}\left(\os{\calt}{-1}_\dm-(\os{A}{-2})^{-1}\os{\calt}{-2}\os{A}{-1}_\dm\right)=0
\end{equation}
Similar invariant combinations can be made for $\os{\calt}{0}_{\dm\dn}$ and the equations, see table \ref{vartab}.

Finally we should point out that in addition to scale and subleading diffeomorphism invariant combinations, the field redefinitions in table \ref{vartab} contain also particular conformal rescalings by powers of $e^\Psi$. Except for the definition of $C_\dm$, where this is related to making the U(1) gauge invariance manifest, these powers in the definitions of the other fields are chosen to simplify the equations. In particular the choice to redefine the spatial metric with such a factor is related to putting the equations in a form that can naturally be obtained from a variational principle.

\subsubsection*{LO equations}
Finally we are ready to present the leading order Einstein and conservation equations. After the redefinitions (\ref{fdef}, \ref{edef}) -- see table \ref{vartab} for details -- one finds that each of $\mathfrak{E}=0\,, \mathfrak{E}_\dm=0\,, \mathfrak{E}_{\dm\dn}=0$ and $\mathfrak{C}=0\,, \mathfrak{C}_\dm=0$ is equivalent to the corresponding equation in the list of equations in table \ref{eqtab}. Note that in those equations all upper indices are indices raised with $k^{\m\n}$ and the connection $\maxnab_\m$ is one preserving the Newton-Cartan structure $(\tau_\m, k^{\m\n})$. There are various such connections, see e.g. \cite{Bekaert:2014bwa,Festuccia:2016awg} for a discussion of the possibilities, but in table \ref{eqtab} we used the connection\footnote{Note that actually any connection of the form $\tilde\maxcon_{\m\n}^\r=\maxcon_{\m\n}^\r+S^\dr\tau_\m\tau_\n+\tau_{\m}S^\dr{}_{\dn}+S^\dr_{\dm}\tau_{\n}$ would leave the equations in table \ref{eqtab} invariant. In that sense \eqref{condef} is the minimal choice. It is related to the boost invariant connections of \cite{VandenBleeken:2017rij, Hansen:2020} by an appropriate choice $S$'s in terms of the higher order fields and a conformal transformation $h^{\m\n}=e^{-\Psi}k^{\m\n}$.} 
\begin{equation}
\maxcon^{\l}_{\m\n} = \tau^\l \partial_\m \tau_\n + \frac{1}{2}k^{\l\r}\left(\partial_\m k_{\r\n} + \partial_\n k_{\r\m} - \partial_\r k_{\m\n}\right)\label{condef}
\end{equation}
Note that this connection has the torsion $\maxcon^{\l}_{[\m\n]}=\tau^\lambda \tau_{[\m}a_{\n]}$. In the equations of \ref{eqtab} also the "Einstein tensor" of this connection appears, 
\begin{equation}
\mathbb{G}_{\dm\dn}=\mathbb{R}_{\dm\dn}-\frac{1}{2}k_{\m\n} k^{\r\s}\mathbb{R}_{\dr\ds}\,,
\end{equation}
as well as the field strength of the gauge potential $C_\dm$\footnote{Note that $C_\mu=C \tau_\m+C_\dm$. The scalar $C$ does actually not appear in \eqref{Fdef} as one can check that $F_{\dm\dn}=h^{\r}_\m h_{\n}^\s(\partial_\r C_\ds-\partial_\s C_\dr)$. Note that this expression is subtly different from $\partial_\dm C_\dn-\partial_\dm C_\dn=h^{\r}_\m \partial_\r C_\dn-h^{\r}_\n\partial_\r C_\dm=F_{\dm\dn}+2\tau_{[\m}h_{\n]}^\r C_\ds\partial_\r\tau^\s$. }
\begin{equation}
F_{\dm\dn}=h^{\r}_\m h_{\n}^\s(\partial_\r C_\s-\partial_\s C_\r)\label{Fdef}\,.
\end{equation}

An important consistency check on the equations in table \ref{eqtab} is that last two equations -- the conservation equations -- follow from the first three equations -- the 'Einstein' equations -- through the Bianchi identity\footnote{See \cite{Hansen:2020} for a comprehensive discussion of various curvature identities of Newton-Cartan connections.} $\maxnab_\dm \mathbb{G}^{\dm\dn}=0$.

\begin{table}[h!]
	\framebox[\linewidth]{\begin{minipage}{0.9\linewidth}
\begin{center}
	{\bf Dynamical variables}
\end{center}
\begin{equation*}
\Psi\,,\quad C_\dm\,,\quad k^{\m\n}\qquad\mbox{and}\qquad \mathfrak{T}\,,\quad\mathfrak{T}_\dm\,,\quad \mathfrak{T}_{\dm\dn}\,.
\end{equation*}
\begin{center}
	{\bf LO Equations}
\end{center}						  
\begin{eqnarray*}
\maxnab^\dr\partial_\dr\Psi &=&\frac{e^{-2\Psi}}{2}F^{\dr\ds}F_{\dr\ds}-16\pi\GN\mathfrak{T}\\
\maxnab_\dr( e^{-2\Psi}F^{\dr\dm})&=&-16\pi\GN\mathfrak{T}^\dm\\
\mathbb{G}^{\dm\dn}&=&\frac{e^{-2\Psi}}{8}\left(k^{\m\n}F^{\dr\ds}F_{\dr\ds}-4F^{\dm\dr}F^{\dn}{}_{\dr}\right)-\frac{1}{4}k^{\m\n}\partial_{\dr}\Psi\partial^{\dr}\Psi\\
&&+\frac{1}{2}\partial^\dm\Psi\partial^\dn\Psi+8\pi\GN\mathfrak{T}^{\dm\dn}\\&&\\
\maxnab_{\dr}\mathfrak{T}^\dr&=&0\\
\maxnab_\dr\mathfrak{T}^{\dr\dm}&=&\mathfrak{T}\partial^\dm \Psi+\mathfrak{T}_\dr F^{\dr\dm}
\end{eqnarray*}
\caption{\label{eqtab} The equations for GR at leading order in the large $c$ expansion.}

\vspace{0.1cm}

\end{minipage}}
\end{table}

\section{Discussion}\label{disc}
We conclude the paper with a number of remarks and observations about the equations in table \ref{eqtab}.
\subsection{Invariance}
Due to the introduction of scale invariant variables -- see table \ref{vartab} -- invariance under the scale symmetry has become trivial. The tensorial nature of the equations guarantees invariance under $c$ independent diffeomorphisms $\os{\xi}{0}$.  Furthermore, in these new variables the subleading diffeomorphisms only act on $C_\dm$ and precisely as a U(1) transformation. The equations are invariant since $C_\dm$ only appears through its gauge invariant curvature $F_{\dm\dn}$. So of the symmetries listed above only Milne boost invariance remains to be considered.

Because some of the variables do transform non-trivially under boosts it might appear as if boost invariance is not manifest. However, only objects with lower indices have a non-zero transformation under the Milne boosts and this will furthermore be proportional to $\tau_\mu$. It follows that all objects that either have all indices raised or contracted with some other raised indices will automatically be boost invariant. As this is exactly the case for the equations in table \ref{eqtab}, these are actually manifestly boost invariant. Let us illustrate this argument with the example of $\partial_\dm\Psi$. By (\ref{tripboost}, \ref{psiboost}) it follows that $\Psi$ is boost invariant so that $\delta_\chi(\partial_\dm\Psi)=\tau_\m\chi^\dr\partial_\dr\Psi$. But then observe that
\begin{equation}
\delta_\chi (\partial^\dm\Psi)=\delta_\chi (k^{\m\r}\partial_\dr\Psi)=k^{\m\r}\delta_\chi (\partial_\dr\Psi)=0\,.
\end{equation}
Similarly $\delta_\chi (\partial^\dm\Psi\partial_\dm\Psi)=\delta_\chi (\maxnab^\dm\partial_\dm\Psi)=0$. 

Note that invariance under Milne boosts as above is possible due to absence of time-like derivatives. At higher order such derivatives will be present. In that case boost invariant variables can be introduced \cite{VandenBleeken:2017rij, Hansen:2020}, but the construction of these variables requires fields that are not present at leading order. 

\subsection{Gauge choices}
As we mentioned in section \ref{setsec}, our approach differs from some of the earlier literature \cite{Dautcourt:1996pm, Tichy:2011te, VandenBleeken:2017rij, Hansen:2020} in that we leave the potential $\os{A}{-2}$ free, rather than choosing it to be $-1$. This does {\it not} amount to the introduction of a new degree of freedom, since upon freeing $\os{A}{-2}$ there appears a local scaling symmetry that in turn removes one scalar degree of freedom. The gauge invariant scalar degree of freedom $\Psi$ -- defined in \eqref{ompart} -- is a combination of $\os{A}{-2}$ and the torsion potential $\psi$ defined in \eqref{apart}. 

\subsubsection*{Dautcourt gauge}
By a scale transformation \eqref{tripscal} one can always make $\os{A}{-2}=-1$, fixing the scaling symmetry. In this choice of gauge our setup reduces to that originally introduced by Dautcourt \cite{Dautcourt:1996pm} and followed in \cite{Tichy:2011te, VandenBleeken:2017rij, Hansen:2020}. In this case the variable $\Psi$ can be identified with the torsion potential $\psi$, or in other words:
\begin{equation}
\mbox{Dautcourt gauge:}\qquad \os{A}{-2}=-1\qquad a_\dm=\frac{1}{2}\partial_{\dm}\Psi
\end{equation}
In this gauge the physical degree of freedom $\Psi$ -- describing a nonrelativistic but strong gravitational time dilation -- finds itself thus in the (twistless) torsion of the Newton-Cartan structure, as was first emphasized in \cite{VandenBleeken:2017rij}.

\subsubsection*{Torsion free gauge}
Alternatively, via \eqref{ascal}, one can also use a scale transformation to put $\psi=0$, again fixing this gauge symmetry. In this gauge the vector $a_\dm$ vanishes and hence the Newton-Cartan structure is torsionless:
\begin{equation}
\mbox{Torsion free gauge:}\qquad \os{A}{-2}=-e^{-\Psi}\qquad a_\dm=0
\end{equation}
This gauge has the advantage that the nonrelativistic geometry used to express the large $c$ expansion is simpler and that the potentials $\os{A}{k}$ are treated equally at all orders. The field redefinitions for the leading order triplet, see table \ref{vartab}, reduce in this gauge to those in table \ref{ourA}.

\begin{table}[!ht]
	\begin{mdframed}
		\begin{center}
			\begin{tabular}{l|ccc}
				LO & $\os{A}{-2}=-e^{-\Psi}$& $\os{A}{-1}_{\dm}=e^{-\Psi}C_{\dm}$ & $\os{A}{0}^{\dm\dn}=e^{\Psi}k^{\m\n}$
			\end{tabular}\caption{The leading order gravitational potentials in torsion free gauge.} \label{ourA}
		\end{center}
	\end{mdframed}
\end{table} 

\subsubsection*{Galilean gauge}
The main motivation behind the Newton-Cartan formalism is to express nonrelativistic gravity in manifestly 4-diffeomorphism invariant form. Still, it turns out that the dynamics restrict $\tau_\mu$ to be twistless \cite{VandenBleeken:2017rij} which means that there is an inherent direction of time in the theory that all observers can agree upon. Furthermore it is clear from table \ref{eqtab}, the equations obtained from the large $c$ expansion take their simplest form in terms of a time vs space split. This all suggests that it might be quite natural to gauge-fix  some of the diffeomorphism invariance by working with an adapted time coordinate. As reviewed in appendix \ref{TTapp}, the twistless torsion condition guarantees that there exists a function $t(x^\mu)$ such that locally $\tau_\mu=e^{-\psi}\partial_\m t$. One can thus choose coordinates $x^\mu=(t,x^i)$ such that $\tau_\mu=e^{-\psi}\delta_\m^0$. Together with a choice of $\tau^\mu=e^{\psi}\delta^\m_0$ these conditions are left invariant by a particular combination of Milne boosts \eqref{NCboost} and time-dependent spatial diffeomorphisms \cite{Bleeken:2015ykr, VandenBleeken:2019gqa}. This gauge condition can be summarized as
\begin{equation}
\mbox{Galilean gauge:}\qquad \tau_\mu=e^{-\psi}\delta_\m^0\,,\quad \tau^\mu=e^{\psi}\delta^\m_0\,,\quad k^{\m0}=k_{\m0}=0\qquad k^{il}k_{lj}=\delta_j^i
\end{equation}
The LO equations in table \ref{eqtab} remain essentially form invariant under this gauge fixing: one simply replaces $\dm\rightarrow i$ and finds that the covariant derivatives and Einstein tensor become those with respect to the Levi-Civita connection of the 3-metric $k_{ij}$. Note that in this gauge the relativistic metric becomes\footnote{See (\ref{m1met}, \ref{0met}) for the error terms.}
\begin{eqnarray}
	ds^2=g_{\m\n}dx^\m dx^\n&=&-e^{-\Psi}(c\, dt+C_i dx^i)^2+e^\Psi k_{ij}dx^idx^j\\&&
	+\calo(c)\, dt^2+\calo_i(c^{0})dtdx^i+\calo_{ij}(c^{-1})dx^idx^j\nonumber
\end{eqnarray}
If one additionally observes that there are no time derivatives in the LO equations in table \eqref{eqtab}, one recognizes these equations as those for an arbitrary stationary relativistic 4-metric. The key crucial difference is that in the large $c$ expansion we performed we did not assume time independence of the fields. But we can conclude that any stationary metric will solve the LO equations exactly. Conversely it follows that at leading order in the large $c$ expansion any solution to the relativistic Einstein equations takes the form of a solution to the stationary Einstein equations but with time-dependent integration constants. We can conclude that the large $c$ expansion is an expansion around the stationary sector of GR. If one makes the coefficients of odd powers vanish -- i.e. take $C_i=0$ above -- the leading order reduces to the static sector, as was already observed in \cite{VandenBleeken:2019gqa}.

Finally we point out that the (vacuum) LO equations, just like those for stationary metrics, can be obtained from a Lagrangian, which follows from a time-like Kaluza-Klein reduction of the Einstein-Hilbert Lagrangian:
\begin{equation}
L=\int d^3x\, \sqrt{k}R-\frac{1}{2}\partial_i\Psi\partial^i \Psi+\frac{e^{-2\Psi}}{4}F_{ij}F^{ij}
\end{equation}

\subsection{Example}
Here we illustrate the expansion procedure and how stationary relativistic metrics provide exact solutions to the LO equations. Starting with the relativistic Kerr metric as in \cite{Wald:1984rg} one finds for the potentials defined through the Newton-Cartan\footnote{We made the choice $\tau_\mu dx^\mu=dt$.} split  \eqref{metricdecompose}
\begin{eqnarray}
A&=& -c^2 \left( 1 - \frac{2\GN mr}{c^2 \S}\right ) \nonumber\\
A_\dm dx^\m &=&-\frac{1}{c}\frac{2a\GN mr\sin^2\theta}{\S} d\phi \label{Aker} \\
A^{\dm\dn}\partial_\m\otimes \partial_\n & =& \frac{\D}{\S} \partial_r^2+\frac{1}{\S}\partial_\theta^2+\frac{\D \csc^2 \theta -a^2  }{\D \S}\partial_\phi^2\nonumber
\end{eqnarray}
where
\begin{equation}
\S = r^2 + a^2 \cos^2\theta\,, \qquad \D =  r^2 + a^2 -\frac{2\GN mr}{c^2 }\,,\qquad a = \frac{J}{mc}\,.
\end{equation}
Depending on how one assumes the mass $m$ and angular momentum $J$ to scale with the speed of light $c$ one gets different expansions.

\subsubsection*{Weakly massive, weakly rotating Kerr metric}
First let us consider the standard Newtonian regime where $\frac{Gm}{r}\ll c^2$ and $\frac{J}{mr}\ll c$, expanding \eqref{Aker} and expressing the fields in terms of the variables of table \ref{vartab} gives
\begin{eqnarray}
\Psi = 0\,,\qquad C_\dm dx^\m= 0\,,\qquad k^{\m\n}\partial_\m\otimes\partial_\n= \delta^{ij}\partial_i\otimes\partial_j
\end{eqnarray}
In this case at leading order the fields simply provide a nonrelativistic description of Minkowski space, the starting point of a weak gravity approximation to GR. The first correction comes in the form of the Newtonian potential $\os{A}{0}=2\Phi=\frac{2GM}{r}$, and then follow further subleading post-Newtonian corrections.

\subsubsection*{Strongly massive, weakly rotating Kerr metric}
Another regime is where $\frac{Gm}{r}\approx c^2$ but $\frac{J}{mr}\ll c$. We can formally implement this regime by defining $m=Mc^2$ and keeping $M$ rather than $m$ fixed as $c\rightarrow \infty$. In this way of expanding the Kerr metric the leading order fields become
\begin{equation*}
\Psi = -\log\left(1- \frac{2\GN M}{r}\right)\,,\qquad C_\dm dx^\m = 0 \,,\ \ 
\end{equation*}
\begin{equation*}
k^{\m\n}\partial_\m\otimes\partial_\n =  \partial_r^2+\frac{1}{r^2-2\GN Mr}\left(\partial_\theta^2+ \csc^2 \theta\partial_\phi^2\right) 
\end{equation*}
In this regime we see that the LO fields contain a spatial metric that is not flat and that translates to a relativistic metric which is not approximately Minkowski. This is an example where the large $c$ expansion extends beyond the regime of weak gravity captured by the post-Minkowski/Newtonian expansion. Note that up to the leading order written here, the Kerr solution coincides with that of Schwarzschild \cite{VandenBleeken:2017rij, Hansen:2020}. Again there is an infinite series of further subleading corrections. Interestingly the Newtonian potential $\os{A}{0}$ vanishes. 

\subsubsection*{Strongly massive, strongly rotating Kerr metric}
The previous expansions of the Kerr solution are free of odd powers of $c$ and as such fall inside the treatment of \cite{VandenBleeken:2017rij, Hansen:2020}. If we however consider a regime where $\frac{Gm}{r}\approx c^2$ and $\frac{Jm}{r}\approx c$ we will see the odd powers appear already at leading order. To set up an expansion around this regime we keep $M=m/c^2$ and $a=J/mc$ fixed as $c\rightarrow\infty$. In this case one finds at leading order
\begin{equation*}
\Psi = -\log \left( 1 - \frac{2\GN Mr}{ \S}\right )\,, \quad
C_\dm dx^\m =-\frac{2a\GN Mr\sin^2 \theta} {\S - 2\GN Mr}d\phi\,,\end{equation*}
\begin{equation*}
k^{\m\n}\partial_\m\otimes\partial_\n  =\frac{1}{\S-2\GN Mr}\left(\D\partial_r^2+\partial_\theta^2+(\csc^2 \theta -\frac{a^2}{\D})\partial_\phi^2\right)
\end{equation*}
where
\begin{equation}
\S = r^2 + a^2 \cos^2\theta\,, \qquad \D =  r^2 + a^2 -2\GN Mr\,.
\end{equation} 
We see here an explicit example of the situation discussed in this paper, namely one where the whole leading order triplet of fields is non-trivial.

Finally let us point out that an analogous weakly massive, strongly rotating regime would be unphysical as it violates the extremality bound.

\subsection{On phenomenological applications}\label{pheno}
The example of the Kerr black hole in the previous subsection shows that there are real world gravitational phenomena that find themselves in the strong field non-relativistic regime that is well approximated by the large $c$ expansion. Still, the expansion is only of real practical use for phenomena that -- contrary to the Kerr black hole -- are too complicated to describe analytically in GR and at the same time require high orders, or a break down, in the post-Minkowski/Newtonian approximation. For such phenomena the large $c$ expansion would an efficient, and possibly unique, analytic tool to compete with numerical GR. What phenomena would qualify as such? The main result of this paper was to identify the large $c$ expansion as an expansion around the stationary sector of GR, and as we pointed out it generalizes the post-Newtonian expansion to include metrics that are not nearly flat. One concludes that possible practical applications of the large $c$ expansion concern phenomena where gravity is strong and that are almost stationary. 

An actual computation of practical value in the large $c$ expansion would hence go at least one order beyond the leading one we worked out, and falls for this reason outside the scope of this paper. Still, to motivate why a further study of the large $c$ expansion might be relevant also to gravitational modeling, we point out two phenomena where an improvement on current techniques might be made.   

\paragraph{Dense rotating stars} In the interior, or close to, massive dense stars, such as neutron stars or white dwarfs, gravity is strong. When the star is rotating at a small, constant angular velocity it is well described by a stationary space-time that approximately solves Einstein's equations, known as the Hartle-Thorn metric \cite{Hartle:1968si}. As we showed in this paper this stationary metric can be taken as a starting point at leading order for the large $c$-expansion and non-stationary corrections can be computed at the next to leading order. This provides an example where one improves on post-Minkowskian/post-Newtonian methods almost by definition, since reproducing the Hartle-Thorn metric itself would already amount to a re-summation of an infinite number of terms of those series.

\paragraph{Black hole binaries} In the current exciting new era of gravitational wave astronomy \cite{Abbott:2016blz} an outstanding question is of course if the large $c$ expansion could help improving models of black hole inspiral. For well separated black holes the initial phase of inspiral is well described by post-Newtonian methods, see e.g.  \cite{Creighton:2011zz} for an introduction. Close to merger, when gravity becomes strong, this expansion reaches its limits and we currently need to rely on numerical methods. The merging of two black holes is of course {\it not} an almost stationary setting either, which suggests that this phenomenon will fall outside the reach of the large $c$-expansion as well. Still, we expect that there exists a regime, at least for certain types of inspiral and for short enough time-scales, where the centers are close enough in orbit that gravity is already strong but motion is not too far from stationarity such that the large $c$-expansion, possibly at a few orders, will give accurate results while post-Newtonian methods would be much less efficient or even break down. It is interesting to recall that stationary solutions describing orbiting black holes have been well studied \cite{Weinstein:1992, Weinstein:2019zrh}. These solutions are typically disqualified as unphysical since they exhibit a conical excess singularity between the holes. It is tempting to speculate however that such a singularity might be an artifact that can be resolved by including non-stationary corrections such as described by the large $c$-expansion. In more basic terms a concrete problem/challenge can be posed as follows. Where the two-body problem in GR is currently out of reach of analytic treatment, its (post-)Newtonian version was solved a long time ago. Can this problem still be solve/addressed analytically in the large $c$-expansion? This expansion has the advantage that it removes radiation, just as the Newtonian version, but gets us closer to GR by including various strong field effects, such as time dilation.

\section*{Acknowledgments}
It is a pleasure to thank E. Bergshoeff, D. Hansen, J. Hartong, N. Obers, M. \"Ozkan, S. Prohazka, J. Raeymaekers and B. Vercnocke for valuable discussions. The work of DVdB is partially supported by the Boğaziçi University Research Fund under grant number 17B03P1 and T\"UBITAK grant 117F376.

\appendix
\section{Some technicalities behind twistless torsion.}\label{TTapp}
A first technical result is that for any two-form $\phi_{\m\n}$ one has the following equivalences
\begin{equation}
\phi_{\dm\dn}=0\qquad\Leftrightarrow\qquad \tau_{[\r}\phi_{\m\n]}=0\qquad\Leftrightarrow\qquad \phi_{\m\n}=2\tau_{[\mu}\phi_{\dn]} \label{equivs}
\end{equation}
The equivalence of the very left and very right follow directly from a decomposition as in section \ref{decompsec}:
\begin{equation}
\phi_{\m\n}=2\tau_{[\m}\phi_{\dn]}+\phi_{\dm\dn}\,,\qquad  \phi_{\dn} =\tau^\r h^\s_\n\phi_{\r\s}\quad \phi_{\dm\dn}=h_\m^\r h^\s_\n\phi_{\r\s}
\end{equation}
Note that the middle equality in \eqref{equivs} is a direct consequence of the equality on the far right. Furthermore observe that via the decomposition above
\begin{equation}
\tau_{[\r}\phi_{\m\n]}=0\quad\Rightarrow\quad \tau_{[\r}\phi_{\dm\dn]}=0  \quad\Rightarrow\quad \tau^\r\tau_{[\r}\phi_{\dm\dn]}=0 \quad\Rightarrow\quad \phi_{\dm\dn}=0
\end{equation}
This then establishes \eqref{equivs}, which by taking $\phi_{\m\n}=\partial_{[\dm}\tau_{\dn]}$ becomes \eqref{tlcond}.

Given that 
\begin{equation}
\partial_{[\m}\tau_{\n]}=\tau_{[\m}a_{\n]}\,,\quad a_\m=a_\dm\label{tlcondapp}
\end{equation}
one can then additionally observe that 
\begin{equation}
	0=\partial_{[\r}\partial_\m \tau_{\n]}
	=-\tau_{[\r}\partial_\m a_{\n]}
\end{equation}
Combining this with \eqref{equivs} for $\phi_{\m\n}=\partial_{[\m}a_{\n]}$ then leads to \eqref{spatialclosed}, which for convenience we reproduce here:
\begin{equation}
\partial_{[\dm}a_{\dn]}=0\label{spatclosapp}
\end{equation}
It follows from a simple calculation that this is satisfied if $a_\mu=\partial_{\dm}\psi$:
\begin{equation}
\partial_{[\dm}a_{\dn]}=h_{[\m}^\r h^\s_{\n]}\partial_\r \partial_{\ds}\psi=\partial_{[\dm} \tau_{\dn]}\tau^\l\partial_\l\psi=0
\end{equation}
We need a more geometric argument to show that \eqref{spatclosapp} actually always implies that (locally) $a_\mu=\partial_{\dm}\psi$. First note that the condition \eqref{tlcondapp} is actually equivalent to the definition of a foliation by hypersurfaces. Consider two purely spatial vectors, i.e. $v^\m=v^\dm$ and $w^\m=w^\dm$, then their commutator will also be purely spatial:
\begin{equation}
\tau_\mu [v,w]^\m=v^\dm w^\dn \partial_{[\m}\tau_{\n]}=0  
\end{equation}
By Frobenius theorem there thus exists a corresponding foliation and, because the kernel of $h^{\m}_\n$ is one dimensional, the leaves are hypersurfaces. We can now argue for $a_\mu=\partial_{\dm}\psi$ in two separate but equivalent ways. One could consider the hypersurfaces to be defined as those surfaces for which the function $t(x)$ is constant. It then follows that because $\tau_\mu v^\m=0$ for any vector $v^\m=v^\dm$ tangent to the hypersurface that
\begin{equation}
\tau_\mu=e^{-\psi}\partial_\m t
\end{equation} 
It then follows that 
\begin{equation}
a_\mu=2\tau^\m \partial_{[\m}\tau_{\n]}=\partial_{\dot \m}\psi
\end{equation}
An alternative route to the same conclusion is to introduce coordinates $y^a$ along a given hypersurface, it follows that
\begin{equation}
\partial_a=e_a^\dm\partial_\dm\label{bas1}
\end{equation}
because the $\partial_a$ form a basis for the hypersurface's tangent space. Furthermore, because a coordinate basis commutes it follows that
\begin{equation}
e_{[a}^\dr\partial_\dr e_{b]}^\dm=0
\end{equation}
Then observe that
\begin{eqnarray}
\partial_{[\dm}a_{\dn]}=0\quad \Rightarrow \quad \partial_{[a}a_{b]}=0
\end{eqnarray}
Via the Poincare lemma on the hypersurface we conclude that
\begin{equation}
a_a=\partial_a\psi\label{loceq}
\end{equation}
Finally we use the fact that \eqref{bas1} is invertible, i.e. since also the $\partial_{\dm}$ form a basis of the tangent space of the hypersurface we can conclude that there exists a matrix $e_a^\dm$ such that $e_a^\dm e^a_\dn=h^\m_\n$. Multiplying both sides of \eqref{loceq} with this matrix we again find $a_{\dm}=\partial_{\dm}\psi$.

\section{From $A$ to $B$}\label{A2B}
The relativistic metric $g_{\m\n}$ and its inverse $g^{\m\n}$ are determined in terms of $(A,A_\dm,A^{\dm\dn})$ through the fields $(B,B^\dm,B_{\dm\dn})$ as in (\ref{metricdecompose}, \ref{Ba}). If one wants to carry out the large $c$ expansion in practice then one will need the expression of the coefficients of the $B$'s in terms of the coefficients of the $A$'s.
Expanding the equations \eqref{Ba} results in the following recurrence relations:
\begin{eqnarray}
\os{B}{k+2}&=&\delta^k_0\oaa^{-1}-\oaa^{-1}\sum_{i=-1}^k\left(\os{A}{i}\os{B}{k-i}+\os{A}{i}_\dm\os{B}{k-i}^\dm\right)\nonumber\\
\os{B}{k+2}^\dm &=&- \oaa^{-1}\sum_{i=-1}^k\left(\os{A}{i}\os{B}{k-i}^\dm+\os{A}{i}_\dn\os{A}{k-i}^{\dn\dm}\right)\label{Brec}\\
\os{B}{k}_{\dm\dn}&=&\delta^k_0 \, h_{\m\n}-\sum_{i=1}^{k+1}\left(\os{A}{k-i}_\dm\os{B}{i}^\dr+\os{B}{k-i}_{\dm\ds}\os{A}{i}^{\ds\dr}\right)h_{\r\n}\nonumber
\end{eqnarray}
To illustrate the use of these recursion relations we work out the first orders of the $B$ fields:
\begin{itemize}
	\item \textbf{LO ($k=-2$): }
	\begin{equation}
	\os{B}{0}=0\qquad \os{B}{0}^\mu=0\qquad \os{B}{-2}_{\dm\dn}=0
	\end{equation}
	Which is equivalent to
	\begin{equation}
	\os{g}{-2}_{\m\n}=\oaa \tau_\mu\tau_\nu\qquad \os{g}{0}^{\m\n}=h^{\m\n}
	\end{equation}
	\item \textbf{N$^{1/2}$LO ($k=-1$): }
	\begin{equation}
	\os{B}{1}=0\qquad \os{B}{1}^\dm=\oaa^{-1}\oa^\dm\qquad \os{B}{-1}_{\dm\dn}=0
	\end{equation}
	Which is equivalent to 
	\begin{eqnarray}
	\os{g}{-1}_{\m\n}&=&\os{A}{-1}\tau_\mu\tau_\nu+\oa_\dm \tau_\nu+\oa_\dn\tau_\mu \label{m1met}\\ 			\os{g}{1}^{\m\n}&=&-\oaa^{-1}(\oa^\dm\tau^\n+\oa^\dn\tau^\mu)+\os{A}{1}^{\dm\dn}
	\end{eqnarray}
	\item \textbf{NLO ($k=0$): }
	\begin{eqnarray}
	\os{B}{2}&=& \oaa^{-2}\oa_\dr\oa^\dr+  \oaa^{-1} \\ 
	\os{B}{2}^\dm &=&\oaa^{-2}\os{A}{-1}\os{A}{-1}^\dm -\oaa^{-1}(\os{A}{1}^{\dm\dn}\oa_\dn+h^{\m\n}\os{A}{0}_\dn )\\
	\os{B}{0}_{\dm\dn}&=& h_{\m\n}+ \oaa^{-1}\oa_\dm\oa_\dn
	\end{eqnarray}
	Which is equivalent to
	\begin{eqnarray}
	\os{g}{0}_{\m\n}&=&\os{A}{0}\tau_\mu\tau_\n+\os{A}{0}_\mu\tau_\nu+\os{A}{0}_\nu\tau_\mu+h_{\m\n}+ \oaa^{-1}\os{A}{-1}_\dm\os{A}{-1}_\dn\label{0met}\\ 
	\os{g}{2}^{\m\n}&=&(\oaa^{-2}\oa_\dr\oa^\dr+  \oaa^{-1} )\tau^\mu \tau^\nu+\oaa^{-2}\tau^\nu\os{A}{-1}_\dr(\os{A}{-1}h^{\r\m}-\oaa\os{A}{1}^{\dr\dm})\nonumber\\
	&&+\oaa^{-2}\tau^\mu\os{A}{-1}_\dr(\os{A}{-1}h^{\r\n}-\oaa\os{A}{1}^{\dr\dn})-\oaa^{-1}(h^{\m\n}+2\os{A}{0}^{(\dm}\tau^{\n)})\nonumber
	\end{eqnarray}
\end{itemize}

\section{A comment on local translations}\label{algapp}
The content of this appendix is somewhat independent of the main text, it served however as an important motivation for the choice of formalism used there. We point out that the link between diffeomorphims and local translations in the nonrelativistic case is more degenerate than in the relativistic case. Where in the relativistic case this degeneracy can be lifted by expressing the Einstein equations in terms of curvatures only, this seems not to be the case in the nonrelativistic setting.

Consider a Lie-algebra valued one form $A$, and a gauge parameter $\Lambda$, which is a zero-form valued in the same algebra. Then we can define the adjoint transformation and curvature as
\begin{equation}
\delta_\mathrm{ad} A=d\Lambda+[A,\Lambda]\qquad F=dA+\frac{1}{2}[A,A]
\end{equation}
The Lie derivative of the gauge field is defined as
\begin{equation}
L_\xi A=d(\ri_\xi A)+\ri_\xi dA
\end{equation}
It follows that
\begin{equation}
L_\xi A=\delta_\mathrm{ad} A+\ri_\xi F\quad\mbox{for }\Lambda=\ri_\xi A\label{difeq}
\end{equation}
Since the Lie derivative generates an infinitesimal diffeomorphism the above equality can be used to translate an adjoint transformation into a diffeomorphism at the cost of an extra curvature contribution. This procedure is very natural if the curvature vanishes by a combination of constraints and dynamic equations, as for example in a frame formulation to general relativity, but less so when this is not the case, as for example in the nonrelativistic approximation to general relativity. Nonetheless this approach remains valid and in \cite{Hansen:2018ofj,Hansen:2020} an algebra was introduced whose translational part reproduces the diffeomorphism symmetries on the gauge field 
\begin{equation}
A = \tau H + m N + e^a P_a + \pi^a T_a + \o^a G_a + \O^a B_a + \frac{1}{2}\o^{ab}J_{ab} + \frac{1}{2}\O^{ab}S_{ab}\,.
\end{equation}
This Lie-algebra is however not the unique one with this feature since in \eqref{difeq} a modification of the adjoint action can be canceled by a modification of the curvature contribution, leading to identical transformations under diffeomorphisms. Demanding that the Lie-algebra is consistent and that the boost and rotational part remains unchanged one can classify all possibilities:
\begin{gather*}
[H,G_a]= P_a  \qquad [N,G_a] = T_a   \qquad [H,B_a] = T_a  \qquad [P_a,G_b] = \delta_{ab}N\\ 
[G_a,G_b] =-S_{ab}\qquad 
[S_{ab},G_c]=-B_a\delta_{bc}+B_b\delta_{ac} \qquad [S_{ab},P_c]=-T_a\delta_{bc}+T_b\delta_{ac}\\ \,
[J_{ab},J_{cd}]=-\delta_{ad}J_{bc}+\delta_{bd}J_{ac}+\delta_{ac}J_{bd}-\delta_{bc}J_{ad}\\ \,
[J_{ab},S_{cd}]=-\delta_{ad}S_{bc}+\delta_{bd}S_{ac}+\delta_{ac}S_{bd}-\delta_{bc}S_{ad}\\ \,
[P_a,P_b]= \alpha S_{ab} \qquad [N,P_a]=\alpha B_a \qquad [H,T_a]= \alpha B_a \\ \qquad [H,P_a] =  \alpha G_a + \beta T_a + \gamma B_a \\
[J_{ab},X_c]=-X_a\delta_{bc}+X_b\delta_{ac}\qquad X_a\in \{P_a,G_a,T_a,B_a\}
\label{algb} 
\end{gather*}
This is a family of algebra's parameterized by the real numbers $\alpha,\beta$ and $\gamma$, that reproduces the algebra of \cite{Hansen:2018ofj, Hansen:2020} when $\alpha=\beta=\gamma=0$. Just as that algebra can be obtained by an expansion procedure from the Poincare algebra \cite{Khasanov:2011jr, Hansen:2020}, the algebras with non-trivial $\alpha$ but $\beta=\gamma=0$ can be obtained by expansion from the (A)dS algebra. For other values of the parameters there doesn't seem to exist any relativistic algebra that they descent from, making them similar to some of the exotic nonrelativistic algebras found in \cite{Figueroa-OFarrill:2019sex}.

\bibliographystyle{utphys}
\bibliography{odd}
\end{document}